\documentclass[11pt,a4paper]{article}                           

\usepackage{amsmath}
\usepackage{amssymb}
\usepackage[usenames]{color} 
\usepackage{multirow}
\usepackage{graphicx}
\usepackage{epstopdf}
\usepackage{array}
\usepackage{hhline}
\usepackage{cite}
\usepackage{microtype}
\usepackage[bf]{caption}
\usepackage{color}
\usepackage[usenames,dvipsnames]{xcolor}

\usepackage{ifpdf}
\ifpdf
  \usepackage[pdftex,
    pdftitle={},
    pdfauthor={},
    pdfsubject={},
    bookmarksopen, bookmarksnumbered, bookmarksopenlevel=2]{hyperref}
\fi
\def\hybrid{\topmargin -20pt    \oddsidemargin 0pt
        \headheight 0pt \headsep 0pt
        \textwidth 6.25in       
        \textheight 9 in       
        \marginparwidth .875in
        \parskip 5pt plus 1pt 
          \jot = 1.5ex
   }
\hybrid
\numberwithin{equation}{section}
\numberwithin{table}{section}

\renewcommand{\phi}{\varphi}
\renewcommand{\theta}{\vartheta}

\newcommand{\p}{\partial}
\newcommand{\rd}{\mathcal}

\newcommand{\be}{\begin{eqnarray}}
\newcommand{\en}{\end{eqnarray}}
\newcommand{\badat}{\begin{alignedat}}
\newcommand{\eadat}{\end{alignedat}}
\newcommand{\bitm}{\begin{itemize}}
\newcommand{\eitm}{\end{itemize}}
\newcommand{\bmat}{\begin{pmatrix}}
\newcommand{\emat}{\end{pmatrix}}
\newcommand{\ba}{\begin{align}}
\newcommand{\bas}{\begin{align*}}
\newcommand{\ab}{\end{align}}
\newcommand{\bse}{\begin{subequations}}
\newcommand{\ese}{\end{subequations}}
\newcommand{\gt}{\rightarrow}

\newcommand{\virg}{\hspace{1 mm}, \hspace{8 mm}}

\def\cL{{\cal L}}

\def\bea{\begin{eqnarray}}
\def\eea{\end{eqnarray}}
\def\ba{\begin{array}}
\def\ea{\end{array}}
\def\bec{\begin{center}}
\def\ec{\end{center}}
\def\ba{\begin{align}}
\def\ena{\end{align}}

\def\12{\frac{1}{2}}

\begin{document}

\baselineskip=14pt
\parskip 5pt plus 1pt

\vspace*{-1.5cm}
\begin{flushright}    
  {\small 
  LMU-ASC 71/17 \\
  MPP-2017-243
  }
\end{flushright}

\vspace{2cm}
\begin{center}        
  {\LARGE Supertranslations and Holography near the  Horizon of 
   \\ 
  \vspace{0.2cm}
  Schwarzschild Black Holes}
\end{center}

\vspace{0.5cm}
\begin{center}        
{\large Dieter L\"ust}
\end{center}

\vspace{0.15cm}
\centerline{\it  Arnold--Sommerfeld--Center for Theoretical Physics,}
\centerline{\it Ludwig--Maximilians--Universit\"at, 80333 M\"unchen, Germany}
\vspace{0.15cm}
\begin{center}        
  \emph{Max-Planck-Institut f\"ur Physik (Werner-Heisenberg-Institut), \\
Fohringer Ring 6, 80805 Munchen, Germany}
             \\[0.15cm]
 
\end{center}

\vspace{2cm}


\begin{abstract}
\noindent
In this paper we review and discuss several aspects of supertranslations and their associated algebras at the horizon of a Schwarzschild black hole.
We will   compare two different approaches
on horizon supertranslations,
which were recently considered in separate publications.
Furthermore we describe a possible 
holographic description of a Schwarzschild black hole in terms of a large N boundary theory, which accommodates the Goldstone bosons of the horizon supertranslations.
\end{abstract}

\thispagestyle{empty}
\clearpage

\setcounter{page}{1}


\newpage

\tableofcontents

\newpage

\section{Introduction}

Over many years, asymptotic symmetries  of certain space-time geometries play an important role in general relativity.
E.g. three-dimensional AdS space  possesses
an infinite-dimensional, asymptotic  $W$-symmetry, which is holographically realized as symmetry group of the two-dimensional conformal field theory that lives on the
boundary of the AdS space \cite{Brown:1986nw}. In four space-time dimensions, the  BMS supertranslations were already introduced in 1962 by Bondi, Metzner and Sachs \cite{bms}.
These infinite-dimensional BMS transformations $BMS^\pm$
describe the  symmetries of asymptotically flat space times at future or past null infinity, denoted by ${\rd I}^+$  and ${\rd I}^-$,
but in general not in the interior of four-dimensional space-time.
Furthermore, if one considers a gravitational scattering process
(an S-matrix in Quantum
Field Theory) on asymptotically flat spaces, there is a non-trivial intertwining between $BMS^+$ and $BMS^-$ (for a recent review on these issues see
\cite{Strominger:2017zoo}).

Recently it was conjectured \cite{Strominger:2013jfa,Hawking:2015qqa}
that the BMS symmetry will play an important role in resolving the so called black hole information paradox by providing an (infinite-dimensional) hair, i.e.
charges 
to the black hole, that carries the information about the collapsing matter before the black hole is formed. In fact, it was argued 
\cite{Hawking:2016msc,Hawking:2016sgy} 
that the BMS group can can be extended
as symmetry group  $BMS^{ H}$ to the horizon ${ H}$ of a Schwarzschild black hole.

In parallel to  the work of \cite{Hawking:2016msc,Hawking:2016sgy},
supertranslations on the horizon of black hole geometries were  considered in a series of papers
 \cite{Dvali:2015rea,Donnay:2015abr,Averin:2016ybl,Averin:2016hhm,DGGP2}. Specifically, in 
\cite{Donnay:2015abr,DGGP2} the supertranslations and superrotations on 
the horizon of four-dimensional metrics as well as  their associated algebras of Killing vectors were constructed. Independently,  the supertranslations  on the horizon 
 of the Schwarz\-schild black hole
were  constructed in \cite{Averin:2016ybl,Averin:2016hhm}.\footnote{Other related work about supertranslations
on horizons can be found in \cite{Banks:2014iha}-\cite{Bousso:2017xyo}.}
 It was shown that the supertranslations on the horizon  generate an infinite number of gapless excitations on the horizon, that potentially can account for the gravitational hair as well as for the black hole entropy.



In this paper we will first review the  construction of the supertranslations  on the event horizon ${ H}$ of a Schwarzschild black hole with metric written in Eddington-Finkelstein coordinates. In particular, one purpose of this paper is to compare the approaches of  \cite{Donnay:2015abr,DGGP2} and of \cite{Dvali:2015rea,Averin:2016ybl,Averin:2016hhm}.
We will find that they are in fact closely related. Furthermore we will show that the supertranslations on the horizon of a Schwarzschild black form an infinite dimensional,
commuting algebra, which is isomorphic to the algebra of supertranslations at null infinity.

In the second part of the paper we will further elaborate on the connection between the BMS transformations and 
the massless Goldstone modes, which are associated to the classical degeneracy of the black hole vacua.
We will discuss a holographic picture of the Schwarzschild black hole, which  is based  on the observations
that the supertranslation algebras on null infinity of asymptotically flat space  and on the black hole horizon agree with each other, and also 
 that flat Rindler space is recovered as
near horizon geometry in the semiclassical large N limit of the black hole geometry. We identify a proper holographic bulk coordinate, denoted by $\lambda(r)$,
which plays the role of a collective coupling constant in the holographic boundary theory. The near horizon limit, where at the horizon $\lambda(r=r_S)=1$, corresponds in the
conformal boundary theory to a quantum critical point, where a Bose-Einstein condensate is formed.
In the 
corpuscular large N quantum picture of the black hole in terms of a 
bound state  of a large number of gravitons    \cite{Nportrait,gold}, a graviton Bose-Einstein condensate is built at $\lambda=1$.


\section{The algebra of supertranslations on the horizon of a  Schwarzschild black hole}

\subsection{Supertranslations  on the horizon}

Here we want to construct the supertranslations on the  event horizon $H$ of a four-dimensional space-time geometry.
First following  \cite{Donnay:2015abr,DGGP2} we write the general near horizon geometry using Gaussian null coordinates, where $v$ represents the retarded time, $\rho \geq 0$ is the radial
distance to the horizon, and $x^A$ ($A=2,3$) are the angular coordinates:

\begin{equation}
ds^2=-2 \kappa \rho \ dv^2+2d\rho dv + 2\Theta_A\rho \ dvdx^A+(\Omega_{AB}+\lambda_{AB}\rho )dx^Adx^B +\Delta g_{ij}\ dx^{i}dx^{j }\,.\label{formA}
\end{equation}
Here $\kappa$ stands for the surface gravity and $\Theta_A$, $\Omega_{AB}$, $\lambda_{AB}$ are arbitrary functions of $v,x^A$, and $\Delta g_{i j}$ being terms of order ${O}(\rho^2)$ ($i,j \in \{v,A\}$). Actually, it is always possible to find a coordinate system in which the metric close to a smooth null surface admits to be written in the form (\ref{formA}) \cite{Moncrief:1983xua, Chrusciel}.

One finds that the asymptotic Killing vector preserving the metric ansatz \eqref{formA} is given by
\begin{equation}
\badat{3}
\label{vectores4D}
&\chi^{v} = f(v,x^A), \\
&\chi^{\rho } =-\p_v f \rho + \frac{1}{2} \Omega^{AB} \Theta_A \partial_{B} f \rho^2 + O(\rho^3),  \\
&\chi^{A} = Y^A(x^B) + \Omega ^{AC}\partial_{C}f  \rho + \frac{1}{2} \Omega^{AD} \Omega^{CB} \lambda_{DB} \p_{C} f \rho^2 + O(\rho^3), 
\eadat
\end{equation}
where $\Omega^{AB}$ is the inverse of $\Omega_{AB}$.

The corresponding variation of the fields read 
\begin{equation}
\badat{3}
&\delta_{\chi} \kappa =Y^{A} \p_{A} \kappa + \p_v (\kappa f) + \p_v^2 f, \label{deltakappa} \\ 
&\delta_{\chi}  \Omega_{AB} = f\partial_v\Omega_{AB} + {\mathcal L}_{Y}\Omega_{AB},\\
&\delta_{\chi} \Theta_A =\cL_Y \Theta_{A}+ f \partial_v\Theta_A -2\kappa \partial_A f -2\partial_v\partial_A f +\Omega^{BD} \partial_v\Omega_{AB} \partial_D f ,\\
&\delta_{\chi}  \lambda_{AB}=f\partial_v \lambda_{AB} -\lambda_{AB}\partial_v f +\cL_Y \lambda_{AB} +\Theta_A\partial_B f+\Theta_B\partial_A f -2\nabla_A\nabla_B f.
\eadat
\end{equation}
By introducing a modified version of Lie brackets \cite{Barnich3}
\begin{equation}
\label{modified}
  [\chi_1,\chi_2]=\mathcal{L}_{\chi_1}\chi_2-\delta_{\chi_1} \chi_2+\delta_{\chi_2} \chi_1 ,
\end{equation}
which suffices to take into account the dependence of the asymptotic Killing vectors upon the functions in the metric, one finds that the algebra of these vectors closes near the
horizon; namely
\begin{equation}
  [\chi(f_1,Y_1^A),\chi(f_2,Y_2^A)]=\chi(f_{12},Y_{12}^A),
	\label{closure}
\end{equation}
where
\begin{equation} \label{modifieddd}
\begin{split}
f_{12}=f_1 \p_{v} f_2-f_2\p_{v} f_1+Y_1^A \p_{A} f_2-Y_2^A\p_{A} f_1,\quad Y_{12}^A=Y_1^B  \p_{B}  Y_2^A -Y_2^B  \p_{B} Y_1^A.  
\end{split}
\end{equation}

\subsection{Schwarzschild black holes}

The set of metrics \eqref{formA} includes in particular the Schwarzschild black hole. 
Namely 
the Schwarzschild-metric in (infalling) Eddington-Finkelstein coordinates $(v,r,\theta,\phi)$ is obtained by the coordinate-transforma\-tion
\begin{equation}
v= t + r* \, ,
\end{equation}
with 
\begin{equation}
dr^* = (1-\frac{r_S}{r})^{-1} dr.
\end{equation}
This choice of Eddington-Finkelstein coordinates covers the exterior and the future-interior of a Schwarzschild-black hole and the Schwarzschild metric gets the form
\begin{equation}\label{Sch}
ds^2 = -(1-\frac{r_S}{r})dv^2 + 2dvdr + r^2 d\Omega^2.
\end{equation} 
 Indeed, performing the change of variable
\be
r=r_S(1+2 \kappa \rho),
\label{map}
\en
where $r_S=1/(2\kappa)$ for Schwarzschild, and expanding in powers of $\rho$, one finds that \eqref{Sch} becomes
\begin{equation}
ds^2=-2 \kappa \rho \ dv^2+2d\rho dv +(\frac{1}{4\kappa^2}+\frac{\rho}{\kappa})(d\theta^2+\sin^2 \theta d\phi^2)+\rd O(\rho^2),
\label{change}
\end{equation}
which belongs to the metric ansatz \eqref{formA} with ($A=\theta,\phi$)
\begin{equation}
\badat{3}
&\Theta_\theta=\Theta_\phi=0,\\
&\Omega_{\theta \theta}=\frac{1}{4\kappa^2} \virg \Omega_{\phi \phi}=\frac{\sin^2 \theta}{4\kappa^2} \virg \Omega_{\theta \phi}=0,\\
&\lambda_{\theta \theta}=\frac{1}{\kappa} \virg \lambda_{\phi \phi}=\frac{\sin^2 \theta}{\kappa} \virg \lambda_{\theta \phi}=0.
\eadat
\end{equation}

Now, let us also restrict the Killing vectors (\ref{vectores4D}) of the BMS transformations on the horizon to the case of the Schwarzschild black hole. 
First, one can check that 
in case of  $\p_v f=0$ and no superrotations $Y^A=0$
 the Killing vectors \eqref{vectores4D}  become:
\begin{equation}
\badat{3}
\zeta^\mu_f&=\left(f(\theta,\phi),0,-4\kappa^2\rho \frac{\p f}{\p \theta},-\frac{4\kappa^2}{\sin^2 \theta} \rho \frac{\p f}{\p \phi}\right)+\rd O(\rho^2),\\
&=\left(f(\theta,\phi),0,-\Omega^{\theta B} \p_B f \rho,-\Omega^{\phi B} \p_B f \rho \right)+\rd O(\rho^2),
\label{vfL2}
\eadat
\end{equation}
Second we use the map \eqref{map}  and
the Killing vectors  on the horizon become
\be
\zeta^\mu_f=\left(f(\theta,\phi),0,\frac{\p f}{\p \theta}(\frac{1}{r}-\frac{1}{r_S}),\frac{1}{\sin^2 \theta}\frac{\p f}{\p \phi}(\frac{1}{r}-\frac{1}{r_S})\right)\, ,
\label{vfL}
\en
where $f(\theta,\phi)$ is an arbitrary function on the two-sphere.

As shown in \cite{Averin:2016ybl,Averin:2016hhm}, these are precisely diffeomorphisms acting on the Schwarzschild metric that leave the horizon invariant. 
According to \cite{Averin:2016ybl,Averin:2016hhm}, the  supertranslations eq.(\ref{vfL}) on the horizon act on the Schwarzschild metric (\ref{Sch}) as follows:
\begin{align}
BMS^H:\quad \delta_{\zeta_f} g_{\mu \nu}
=
\begin{pmatrix}
0 & 0 & -(1-\frac{r_S}{r})\frac{\partial f}{\partial \theta} & -(1-\frac{r_S}{r})\frac{\partial f}{\partial \phi}\\
0 & 0 & 0 & 0\\
* & 0 & 2r^2(\frac{1}{r}-\frac{1}{r_S})\frac{\partial^2 f}{\partial \theta^2} & 2r^2(\frac{1}{r}-\frac{1}{r_S})(\frac{\partial^2 f}{\partial \theta \partial \phi}-\cot \theta \frac{\partial f}{\partial \phi})\\
* & 0 & * & 2r^2(\frac{1}{r}-\frac{1}{r_S})(\frac{\partial^2 f}{\partial  \phi^2} + \sin \theta \cos \theta \frac{\partial f}{\partial \theta})
\end{pmatrix}
.
\label{Bogoliubov}
\end{align}

Using complex coordinates\footnote{The complex coordinates are defined as $z=\cot\bigl({1\over 2}\theta\bigr)e^{i\phi}$
and the metric of $S^2$ in angular coordinates reads $ds^2=2r^2\gamma_{z\bar z}dzd\bar z=d\theta^2+\sin^2\theta d\phi^2$.}
on the two-sphere the Killing vectors (\ref{vfL}) can be expressed as:
\begin{equation}\label{killingh}
BMS^H:\quad \,\quad \zeta_f=f{\partial\over\partial v}-{r_S-r\over rr_S}(D^zf{\partial\over \partial z}+h.c.)\, .
\end{equation}
They act on the coordinates $(v,r,z,\bar z)$ in the following form:
\begin{eqnarray}\label{horizontrans}
 BMS^H:\quad\,\quad v&\rightarrow& v-f(z,\bar z)\, ,\nonumber\\
z&\rightarrow &z+{r_S-r\over rr_S}\gamma^{z\bar z}\partial_{\bar z}f(z,\bar z)\, ,\nonumber\\
\bar z&\rightarrow &\bar z+{r_S-r\over rr_S}\gamma^{z\bar z}\partial_{ z}f(z,\bar z)\, ,\nonumber\\
r&\rightarrow &r\, .
\end{eqnarray}

Now we can also determine the algebra of the Killing vectors of the Schwarz\-schild black hole. Namely we want to
compute the algebra formed by the vector fields \eqref{vfL} and see if would reproduce a BMS algebra. Taking the commutator of two vector fields, one finds 
near to the horizon, i.e. $r \gt r_S$, that
the algebra  closes  and becomes an infinite-dimensional, commutative supertranslation algebra: 
\be
[\zeta_{f_1},\zeta_{f_2}]_{r \gt r_S}&=&\left(f_{12}
+\rd O(\rho^{2}),\rd O(\rho^{2}),\rd O(\rho^{2}),\rd O(\rho^{2})\right)\nonumber\\
&=&\left(\rd O(\rho^{2}),\rd O(\rho^{2}),\rd O(\rho^{2}),\rd O(\rho^{2})\right).
\label{algh}
\en
This is consistent with the fact, when going to the Schwarzschild metric, we have considered only those supertranslations satisfying $\p_v f=0$ and with no superrotation $Y^A=0$.
It then follows that $f_{12}=0$.

We can also consider the standard BMS-supertranslations of the Schwarz\-schild metric on   ${\rd I}^-$. These are simply obtained by 
taking the limit $r_S \to \infty$     \cite{Dvali:2015rea,Averin:2016ybl}, which takes the future horizon $r=r_S$ to past null infinity ${\rd I}^-$.  
In this limit the vector fields reduce to
\be
\zeta^\mu_f=\left(f(\theta,\phi),0,\frac{\p f}{\p \theta}\frac{1}{r},\frac{1}{\sin^2 \theta}\frac{\p f}{\p \phi}\frac{1}{r})\right),
\en
which is a BMS supertranslation vector field at null infinity  with respect to the Schwarz\-schild metric.\footnote{Notice though that $\xi^r=\frac{1}{2}\Delta f$ for BMS.}
Like the algebra on the horizon in eq.(\ref{algh}),  
  one gets in this case that 
\be
[\zeta_{f_1},\zeta_{f_2}]_{r_S \gt \infty}=\left(\rd O(r^{-2}),\rd O(r^{-2}),\rd O(r^{-2}),\rd O(r^{-2})\right),
\label{algb}
\en
and hence the algebra closes with $f_{12}=0$, namely one recovers the usual supertranslation algebra
\be
[f_1,f_2]=0.
\en

\vskip0.5cm

As discussed in \cite{Averin:2016ybl,Averin:2016hhm}, the BMS transformations on the horizon also contain in part BMS transformations on $\rd I^-$. Since 
the BMS transformations
on null infinity are not supposed to contribute to the black hole entropy,  we want to subtract them from the supertranslations
on the horizon in order to isolate the physical gapless modes of the black hole.  This subtraction was performed \cite{Averin:2016ybl} for the infinitesimal
supertranslations, and we have called the modes that result from this factorization the ${\cal A}$-modes on the horizon.
Furthermore, 
as one can see,  the modes of $BMS^H$ in eq.(\ref{Bogoliubov})  have also non-vanishing entries in the $v-\theta$ and  $v-\phi$ components of these matrices. 
These are kind of {\sl electric} field strength components.
We want to 
construct transformations where
these components are however absent. This is precisely also the case 
 for the $\cal A$-modes. 
  Namely the effect of  taking the quotient
$\cal A$ erases the electric components of the gapless modes and leave only the magnetic components.
These transformations act on the Schwarzschild metric eq.(\ref{Sch}) in the following way:
\begin{align}
{\cal A}:\quad\delta_{\chi_f}g_{\mu \nu}=
\begin{pmatrix}
0 & 0 & 0 & 0\\
0 & 0 & 0 & 0\\
0& 0 & -2r^2 \frac{1}{r_S} \frac{\partial^2 f}{\partial \theta^2} & -2r^2 \frac{1}{r_S} (\frac{\partial^2 f}{\partial \theta \partial \phi}-\cot \theta \frac{\partial f}{\partial \phi})\\
0 & 0 & * & -2r^2 \frac{1}{r_S} (\frac{\partial^2 f}{\partial  \phi^2} + \sin \theta \cos \theta \frac{\partial f}{\partial \theta}).
\end{pmatrix}
\label{microstates}
\end{align}  

The corresponding modes constitute an infinite family of classical black hole metrics. They are the physical, classical Goldstone modes of the supertranslations at the horizon.
We see that the electric components of the metric fluctuations are indeed absent in for the $\cal A$-modes. So the $\cal A$-modes describe soft  gravitational waves that are confined to the horizon and do not fluctuate into the bulk.
The $\cal A$-modes  feel small changes of $r_S$
only with respect to their amplitudes, as can be seen in above equation. 
 Hence the
$\cal A$-modes constitute a kind of {\sl membrane paradigm}. 
One can in addition show that 
 the metric $g_{\mu\nu}^f=g_{\mu\nu}+\delta_{\chi_f}g_{\mu \nu}$  is Ricci-flat  in the angular directions:
 \begin{equation}
 R_{\theta\phi}(g_{\mu\nu}^f)=R_{\theta\theta}(g_{\mu\nu}^f)=R_{\phi\phi}(g_{\mu\nu}^f)=0\, .
 \end{equation}
 Hence the Goldstone modes along the horizon fully solve the non-linear vacuum Einstein equations in the angular directions. Hence no energy momentum tensor in the
 angular directors needs to be generated. 

\vskip0.2cm

\section{Family of Schwarzschild metrics and soft modes on the horizon}

In this section we will continue to discuss the family of Schwarzschild metrics, which are related to the supertranslations on the horizon, in more detail, where we will use a slightly
different form of the metric compared to previous section.

\subsection{Family of flat metrics and soft modes at infinity}

Let us first briefly recall the expansion of  the metric in Bondi coordinates at around $\rd I^-$ to  
obtain a family of asymptotically flat metrics (see \cite{Strominger:2017zoo}  for a review on that subject):
\begin{eqnarray}\label{bondimetric}
ds^2&=&-dv^2+ dvdr+2r^2\gamma_{z\bar z}dzd\bar z\nonumber\\
&+&{2m_B\over r}dv^2+rC_{zz} dz^2+rC_{\bar z\bar z}d\bar z^2-2U_zdvdz-2U_{\bar z}dvd\bar z+\dots \, .
\end{eqnarray}
Here $\gamma_{z\bar z}={2\over(1+z\bar z)^2}$ is the round metric  of the unit $S^2$ and
\begin{equation}
U_z=-{1\over 2}D^zC_{zz}\, .
\end{equation}
Furthermore, $m_B$ is the Bondi mass for gravitational radiation, the $C_{zz}(v,z,\bar z)$ are in general functions of $z,\bar z,u$  and the Bondi news $N_{zz}(v,z,\bar z)$ are characterizing ingoing gravitational waves that pass trough $\rd I^-$:
\begin{equation}
N_{zz}(v,z,\bar z)=\partial_uC_{zz}(v,z,\bar z)\, .
\end{equation}
Gravitational vacua with zero radiation have $N_{zz}=0$.
In gauge theory language, the metric component $C_{zz}(v,z,\bar z)$ plays the role of the gauge connection, whereas the Bondi news $N_{zz}(v,z,\bar z)$ are analogous to the field strength in gauge theory.
One can show that the standard supertranslations on $\rd I^-$ with some function $g(v,z,\bar z)$  act on the Bondi mass and on 
gauge connection $C_{zz}$ as:
\begin{equation}\label{ctrans}
m_B\,\rightarrow \, m_B\, ,\quad C_{zz}\,\rightarrow \, C_{zz}-2D^2_zg\, .
\end{equation}
The term $2D^2_zg$ in above equation is a pure gauge and does not led to any field strength $N_{zz}=0$.
It follows that the field strength, i.e. the Bondi news are invariant under BMS transformations:
\begin{equation}\label{ctrans}
N_{zz}\,\rightarrow \, N_{zz}\, .
\end{equation}
Hence, for a static background with zero radiation, namely if $N_{zz}=0$  like for Minkowski space, BMS transformation do not generate any Bondi news.

Nevertheless static backgrounds, as Minkowski space, are not unique, but they are characterized by an infinite family of pure gauge connections $ C_{zz}=2D^2_zg$.
It then follows that an infinite family of  flat metrics can be defined as those satisfying 
\begin{equation}
 C_{zz}=D^2_z\Phi(z,\bar z)\, ,
 \end{equation}
 with $\Phi(z,\bar z)$ being some arbitrary function.  The associated Bondi news are automatically zero. In other words, we can define the manifold ${\cal C}$ of flat vacua as 
 \begin{equation}
 {\cal C}=\large\lbrace  C_{zz}\, ,\, C_{zz}=D^2_z\Phi(z,\bar z)\large\rbrace\,.
 \end{equation}
Hence  the space of flat vacua is isomorphic to the space of functions $\Phi(z,\bar z)$ on the 2-sphere. BMS transformations are transforming points of ${\cal C}$ onto points of ${\cal C}$:
 \begin{equation}
 BMS:\quad C_{zz}=D^2_z\Phi\,\longrightarrow C^g_{zz}=D^2_z(\Phi-2g)
 \,.
 \end{equation}




We can translate this discussion  into a quantum mechanical picture: the space of classical (asymptotically)  flat metrics corresponds
to family of classical vacua $|\Phi\rangle$ as follows:
\begin{equation}
C_{zz}=D^2_z\Phi\,\longrightarrow |\Phi\rangle\equiv |C_{zz}\rangle=|D^2_z\Phi\rangle\, .
\end{equation}
Furthermore BMS supertranslation generators are given in terms of charges (generators) $Q^-$ and act on the classical vacua as
\begin{equation}
Q^-|\Phi\rangle=|C^g_{zz}\rangle=|D^2_z(\Phi-2g)\rangle\equiv |\Phi-2g\rangle\, .
\end{equation}
Since the $Q^-$ do not annihilate the ground state $|\Phi\rangle$,  they  are the generators of the spontaneously broken BMS symmetry.

Next let us discuss the 
Goldstone modes, which are associated to the spontaneously broken generators $Q^-$.
Classically, the  Goldstone modes  just correspond to the variations of the classical metric   under the BMS supertranslations, denoted by $\delta_gg_{\mu\nu}$.
Since the Goldstone modes will correspond to physical zero momentum gravitons in the quantum theory, we want to focus on the physical, transversal metric fluctuation, namely
on the $\delta_gg_{zz}$ component of the metric fluctuation 
 (and similarly also $\delta_gg_{\bar z\bar z})$. According to the discussion in the previous section, the Goldstone modes  are then nothing else than  $\delta_gC_{zz}=2D^2_zg$.

Now we are at a stage to  relate the Goldstone modes to the soft gravitons. For this purpose we are considering two classical vacua at some initial time $v_i$ and some
final time $v_f$, where the two vacua differ from each other by a BMS transformation:
\begin{equation}
v_i:\quad C_{zz}^i=D^2_z\Phi_i\, ,\qquad v_f:\quad C_{zz}^f=D^2_z\Phi_f=D^2_z(\Phi_i-2g)\, .
\end{equation}
In addition we are considering an interpolating metric $C^{int}_{zz}(u,z,\bar z)$ with the properties 
\begin{equation}
C^{int}_{zz}(v_i,z,\bar z)= C_{zz}^i\, ,\qquad  C^{int}_{zz}(v_f,z,\bar z)= C_{zz}^f\, .
\end{equation}
The interpolating metric can be regarded as a domain wall that interpolates between these two classical vacua. It necessarily possesses non-vanishing Bondi news:
\begin{equation}
N_{zz}=\partial_vC^{int}_{zz}(v,z,\bar z)\neq 0\, .
\end{equation}
Hence the domain wall corresponds to a time-dependent 
radiative mode with non-zero Bondi news, which describes a gravitational wave that passes $\rd I^-$ between the times $v_i$ and $v_f$. 
The zero frequency part of this gravitational wave is called {\sl soft graviton} mode and is defined in the following way:
\begin{equation}
N_{zz}^{\omega=0}=\lim_{\omega\rightarrow 0}\int_{v_i}^{v_f}dv~e^{i\omega v}\partial_vC_{zz}(v,z,\bar z)\, .
\end{equation}
Using partial integration it is simply given by the BMS variation of the metric component $C_{zz}$:
\begin{equation}
N_{zz}^{\omega=0}=C_{zz}^f-C_{zz}^i=\delta_gC_{zz}=2D^2_zg\,.
\end{equation}
This shows that the 
Goldstone mode is just the soft graviton:
\begin{equation}
|{\rm Goldstone~boson}\rangle \equiv |{\rm Soft ~graviton}\rangle\, .
\end{equation}
It is interesting to note that even Minkowski space is equipped with low energy Goldstone modes. However as we will discuss now, the corresponding BMS charges
of Minkowski space are zero.

So let us finally construct the corresponding BMS charges. In general the classical Noether charge is defined as
\begin{equation}
Q^-=\omega(C_{zz},{\cal L}_gC_{zz})
\, ,
\end{equation}
where $\omega$ is the symplectic form associated to the classical phase space. In our case we obtain
\begin{eqnarray}
Q^-&\simeq&\int_{\cal C}dv~ dz~d\bar z~\partial_vC_{zz}D^2_zg=\nonumber\\
&=&-\int_{\cal C}dv~ dz~d\bar z~C_{zz}\partial_v(D^2_zg)=\nonumber\\
&=&-{1\over 2}\int_{\cal C}dv~ dz~d\bar z~C_{zz}\partial_v(\delta_gC_{zz})\, .\label{BMScharges}
\end{eqnarray}
where the Cauchy surface ${\cal C}$ is null and can be identified with $I^-$.
The BMS charges also relate to the soft gravitons, i.e. to the Goldstone bosons via the standard low energy theorem:
\begin{equation}
\langle0|Q^-|GB\rangle\neq0
\end{equation}

However it is important to note that eq.(\ref{BMScharges}) implies that 
the BMS charges are zero for static, i.e. $v$-independent backgrounds. Therefore the BMS charges for Minkowski space are vanishing.
The  
 BMS charges can only be non-zero, if the background itself is $v$-dependent and
describes some radiative modes. In other words, if a soft gravitational wave is excited, which has the form of a gravitational plus at $\rd I^-$, the BMS charges are non-zero.

The vanishing of the BMS charges  for static configurations can be also understood using the quantum language. Consider 
the expectation value of the charge operator between a vacuum state $|\Phi\rangle$:
\begin{equation}
\langle\Phi|Q^-|\Phi\rangle=\langle\Phi|\Phi-2g\rangle
\end{equation}
This expectation value vanishes for static vacua, since in this case $ |\Phi\rangle$   and $|\Phi-2g\rangle  $ are orthogonal to each other.


\subsection{Family of black holes under horizon supertranslations:}

We now like to derive the soft modes, which are associated to the supertranslations in the horizon, in a similar fashion to the ones at $\rd I^-$.
Therefore 
 we  now  expand the Schwarzschild metric around  in $({r}-{r_s})^{-1}$ around  the horizon in the following form:
\begin{eqnarray}
ds^2 &=& -(1-\frac{r_S}{r})dv^2 + 2dvdr + r^2 \gamma_{z\bar z}dzd\bar z\nonumber\\
&+& {rr_S\over r-r_S}C^H_{zz} dz^2+{rr_S\over r-r_S}C^H_{\bar z\bar z}d\bar z^2+\dots \, .
\,.
\end{eqnarray} 
The  $C^{H}_{zz}(v,z,\bar z) $ play again the role of gauge connections on the horizon.
Non-vanishing radiation at the horizon is now characterized by a new kind of Bondi news which are the associated field strengths and take the form
\begin{equation}
N^{H}_{zz}=\partial_vC^{H}_{zz}(v,z,\bar z)\, .
\end{equation}
For static backgrounds, like the Schwarzschild metric, the $C^{H}_{zz}$ do not depend on $v$ and the classical radiation $N^{H}_{zz}$ is zero.

In eqs.(\ref{Bogoliubov})  we have written the transformation of the metric under $BMS^H$ supertranslations in matrix form.
We can now equivalently express these transformations as transformations on 
the
gauge connections $C_{zz}^{H}$ as:  
\begin{equation}\label{cHtrans}
 BMS^H:\quad C^{H}_{zz}\,\rightarrow \, C^{H}_{zz}-2D^2_zf\, .
\end{equation}

We  see  that the supertranslations on the horizon generate an infinite family of static black hole metrics, all characterized by pure  gauge connections $C^{H}_{zz}$,
namely those satisfying
\begin{equation}
 C^{H}_{zz}=D^2_z\Phi^{H}(z,\bar z)\, ,
 \end{equation}
 with  now $\Phi^{H}(z,\bar z)$ being some arbitrary function.   So, we can define the manifold ${\cal C}^{H}$  of Schwarzschild vacua as 
 \begin{equation}
 {\cal C}^{H}=\large\lbrace  C^{H}_{zz}\, ,\, C^{H}_{zz}=D^2_z\Phi^{H}(z,\bar z)\large\rbrace\,.
 \end{equation}
Hence  the space of Schwarzschild vacua is isomorphic to the space of functions $\Phi^{H}(z,\bar z)$ on the 2-sphere.  
$BMS^H$ supertranslations on the horizon are transforming points of ${\cal C}^{H}$ onto points of ${\cal C}^{H}$:
 \begin{equation}
 BMS^H:\quad C^{H}_{zz}=D^2_z\Phi^{H}\,\longrightarrow C^{H,f}_{zz}=D^2_z(\Phi^{H}-2f)
 \,.
 \end{equation}

Now we come to the charges and to the soft modes on the horizon related to $BMS^H$. The discussion is 
quite analogous to the one in the previous section, and
hence we will not provide all details. 
Namely the corresponding soft graviton on the horizon is  defined in the following way:
\begin{equation}
N_{zz}^{H,\omega=0}=\lim_{\omega\rightarrow 0}\int_{v_i}^{v_f}dv~e^{i\omega v}\partial_vC^{H}_{zz}(v,z,\bar z)=\delta_fC^{H}_{zz}=2D^2_zf\, .
\end{equation}
In addition   we then get the following expression for the supertranslation charges at the horizon
\begin{eqnarray}
Q^{H}&=&\int_{I_{B.H}}dv~ dz~d\bar z~\partial_vC^{H}_{zz}D^2_zf=\nonumber\\
&=&-\int_{I_{B.H.}}dv~ dz~d\bar z~C^{H}_{zz}\partial_v(D^2_zf)=\nonumber\\
&=&-{1\over 2}\int_{I_{B.H.}}dv~ dz~d\bar z~C^{H}_{zz}\partial_v(\delta_fC^{H}_{zz})\, .\label{BMSAcharges}
\end{eqnarray}
It is clear that for the eternal static black hole these charges are zero, because already the integrand is vanishing, since the $H$-modes are $v$-independent

\vskip1cm

At the end of this section we like to emphasize that for the classical eternal black hole, we are discussing 
at the moment, the classical Goldstone modes of the  H- and ${\cal A}$-supertranslations do not carry any energy, i.e. they have zero frequency.
In other words, for the eternal, static black hole no energy is deposited  on the horizon by the classical $BMS^{H/{\cal A}}$ supertranslations. 
The fact that the modes have zero frequency is also the reason why the corresponding classical charges are zero for the eternal Schwarzschild black hole.
On the other hand, for classical backgrounds, which are not static, like collapsing shells, the $H/\cal A$-modes will in general  possess non-zero frequencies. This is the
case if gravitons or also other particles with gravitational interactions fall into the black hole. In this case 
the  corresponding charges are possibly non-zero.
So  the BMS charges or the $H/{\cal A}$-charges do not only measure the number of how many soft gravitons are at the horizon, but each charge also measures 
the frequency of these modes.

In summary, the classical information, which is associated to the infinite classical entropy of a static black hole,  is not accessible by any classical experiment. In other words, the infinite classical
entropy, i.e. the infinite classical hair cannot be resolved in a finite amount of time - see also \cite{Bousso:2017xyo}.
 Also the classical electromagnetic hair considered in \cite{Hawking:2016msc} also would
need infinite time to be resolved.

\section{Schwarzschild holography and the quantum black hole case}

\subsection{Near horizon limit of the Schwarzschild geometry and  a holographic bulk coordinate $\lambda$}

It is known from the work of Maldacena \cite{Maldacena:1997re}
 that the near horizon limit of $N$ coincident D3-branes for  $N\rightarrow\infty$ is described by the 
product space $AdS_5\times S^5$. The AdS/CFT correspondence then states that type IIB supergravity on $AdS_5\times S^5$ is holographically equivalent to ${\cal N}=4$
super Yang-Mills gauge theory with gauge group $SU(N)$ on the four-dimensional boundary of $AdS_5$. In the supergravity limit, i.e. in the limit of small
$AdS_5$ curvature radius,  besides $N$ also the t'Hooft coupling  $\lambda=g_s N$ of the $SU(N)$ gauge theory  is taken to be large.

In section two we have derived the infinite-dimensional commuting algebra of the supertranslations on the Schwarzschild horizon, which is the same as the usual algebra of supertranslations at infinity. 
At first sight it looks surprising that the two supertranslation algebras are the same.  But this becomes understandable by noting that 
the near-horizon geometry of the Schwarzschild black hole is given by the flat Rindler space.
Motivated by this observation, 
 we want to discuss in the following a  picture, which is in close analogy to the emergence of $AdS$ geometry via the  branes in superstring theory:
 for the Schwarzschild geometry
 there exist a three-dimensional  
  dual conformal theory  (reps. one-dimensional theory, when omitting the angular coordinates)
  living on the holographic screen of Rindler space, namely the one-dimensional Rindler cone of $M^{1,1}(\times S^2)$, in analogy to the $SU(N)$ gauge theory on the boundary of $AdS_5(\times S^5)$.



First recall that classical Bekenstein-Hawking entropy   ${\cal S}$   \cite{Bek,Hawking}, here also denoted by $N$,
 of a Schwarz\-schild black hole, the radius $r_S$ of the horizon and the black hole mass $M$ scale scale   as 
    \begin{equation}
 {\cal S}=N\, ,\quad r_S=L_P\sqrt N\, ,\quad M=M_P\sqrt N \, .
 \label{N}
 \end{equation}
  Here 
  $L_P$ ($M_P$) is Planck length (mass), which in terms of $G_N$ and Planck's constant $\hbar$ is defined as 
   $  
 L_P^2 \, \equiv \, \hbar G_N $.
As usual,  the gravitational radius  and the mass of the black hole are related as $r_S = G_NM$.

As just said,
the near-horizon geometry of the Schwarzschild geometry is given by the flat Rindler space, denoted by 
$M^{1,1} \times S^2$, where $M^{1,1}$ is the 2-dimensional Minkowski space in Rindler coordinates $\tilde t$ and $\tilde\rho$. Its metric is given as 
  \begin{equation}
 ds^2=-\tilde\rho^2d\tilde t^2+d\tilde\rho^2+\sum_{i=2,3}dx^idx_i\,. 
 \end{equation}
The Rindler coordinates are related to the Schwarzschild coordinates as follows
\begin{equation}
\tilde \rho=\sqrt{r_S(r-r_S)}=\sqrt{r_S\rho}\, ,\qquad \tilde t={t\over r_S}\, .
 \end{equation}
 In the following we consider 
  for fixed $\hbar$ and $G_N$, the following semiclassical large $N$ limit,  such that $\tilde\rho$ and $\tilde t$ are finite for small $\rho$ reps. for large $t$:
  \begin{equation}\label{largeN}
N\rightarrow\infty\, ,\quad r_S\rightarrow\infty\, , \quad M\rightarrow\infty\,,
\end{equation}
So 
in this limit  the entropy as well as the mass and the horizon become infinitely large.

We now want to consider the Schwarzschild horizon is a kind of 3-dimensio\-nal holographic boundary with coordinates $t$, $\theta$ and $\phi$ or respectively in the
near horizon Rindler geometry with boundary coordinates $\tilde t$, $x_2$ and $x_3$. The orthogonal bulk coordinate is given by $\rho$ or by $\tilde\rho$,
respectively. 
 The bulk coordinate $\rho$ (or $\tilde \rho$) can be related to a new holographic bulk coordinate $\lambda$ in the following way:
\begin{equation}
\lambda(r) = {L_P^2\over r^2} N={r_S^2\over r^2}={r_S^2\over (\rho+r_S)^2}=1-{\rho\over 2r_S}+\dots
\end{equation}
The horizon is now located at the point $\lambda(r=r_s)=1$.
So the near-horizon limit $\rho\rightarrow 0$ corresponds to the limit
\begin{equation}
\lambda\rightarrow 1\,.
\end{equation}
Note that $\lambda$ can be rewritten as 
\begin{equation}
\lambda=       \alpha N ,
\end{equation}
where 
\begin{equation}
\alpha(r)= {L_P^2\over r^2}={r_S^2\over r^2}{1\over N}
\end{equation}
 is nothing else than the scale dependent gravitational coupling constant. It becomes very small in the large N limit.
In fact, the coupling constant $\lambda$ 
was introduced in    
\cite{Nportrait,gold}
 as collective coupling constant of  N gravitons that form a black hole boundstate, where the coupling constant between two individual graviton pairs is just given by $\alpha$.
 It was further argued that in the near horizon limit, namely at the point $\lambda=1$, the N graviton system becomes a quantum critical system, which behaves like a Bose-Einstein condensate at
 the quantum critical point.

\subsection{Holographic bulk-boundary correspondence}

In the following we want to further  explore the idea that quantum gravity in the four-dimensional Schwarzschild bulk geometry is dual to a quantum field theory with 
lives on a three-dimensional boundary, namely the black hole horizon $H$. So what are the expected features of such a holographic bulk-boundary correspondence?

Let us first discuss some qualitative properties of the holographic Schwarz\-schild bulk-boundary correspondence
in the large N limit eq.(\ref{largeN}), where the black hole horizon becomes infinite, and the gravity theory becomes semiclassical. 
The large N limit means
that the  coupling constant $\alpha$ of the three-dimensional boundary field theory becomes very weak.
Furthermore at the near horizon limit
 $\lambda=1$ in the bulk, the boundary theory should become gapless \cite{Nportrait,gold}. This is the point of quantum criticality of
the boundary theory, where its collective coupling $\lambda=\alpha N$
 becomes critical, i.e. $\lambda=1$. Then
 the spectrum of the three-dimensional boundary theory will become highly degenerate with an infinite number of massless states that account for
 the infinite black hole entropy. Furthermore the boundary theory should become conformal.
 On the other hand, for a finite black hole horizon, i.e. finite N, the bulk gravity theory will receive $1/N$ quantum corrections.
 For   the boundary theory this means that the energy gap will disappear, where the splitting of the energy levels will be of order $1/N$.
 These $1/N$ corrections will also destroy the conformality of the boundary theory.

 In summary we expect that in the large $N$, near horizon limit, semiclassical four-dimensional gravity of the Schwarzschild black hole can be described by a 
 highly degenerate, three-dimensional
 gapless theory, which will be a particular three-dimensional 
 conformal field theory.

As we have seen so far, for classical gravity the supertranslation charges on the horizon are vanishing.
 In the quantum language, the  Goldstone modes, also now called Bogoliubov modes, of the classical supertranslations in eqs.(\ref{Bogoliubov},\ref{microstates}) and eq.(\ref{cHtrans})
correspond  to the Goldstone bosons of the spontaneously broken supertranslations at the horizon:
\begin{equation}
|{\rm Goldstone~boson}\rangle \equiv |{\rm Bogoliubov ~mode}\rangle\, .
\end{equation}
As we will now explain, these Goldstone bosons are composed out of gravitons, however unlike to the  soft gravitons at $\rd I^-$, they are not  asymptotic, 
perturbative  spin-two graviton particles, but rather they are collective modes of gravitons. Furthermore, in the quantum theory with finite $N$, the Goldstone modes are not completely
gapless (massless) anymore, but they acquire a mass, which however is still small compared to the black hole mass itself.

So in the quantum case the situation compared to the classical situation substantially changes:
The 
black hole is not anymore eternal, it radiates and the radiation also implies a back reaction on the classical metric, such that it is not any more static. 
Furthermore due the finite size of the horizon, quantum mechanics imply that the  $H/{\cal A}$-modes are described by Bogoliubov modes, which are not any more gapless and do not 
correspond to  zero-momentum gravitons, but to
modes on the horizon with now finite frequency, i.e. with finite energy.
In other words, in the quantum theory there is a finite energy gap, denoted by the frequencies $\omega$, which are of order $1/N$.
It follows that in the quantum case, the supertranslations  on the horizon are explicitly broken, which means that the Bogoliubov modes are not anymore massless Goldstone
bosons, but will acquire a mass of the order of the energy gap.

As we will now indicate, it furthermore  follows that the quantum  supertranslation charges on the horizon will be non-vanishing.
The corresponding entropy also becomes finite, where it was formally infinite in the classical case.
We can define the quantum charge  operator $\hat {\cal Q}_{lm}^{H/{\cal A}}$  for
the supertranslations at the horizon in the same way as the classical charge in eq.(\ref{BMSAcharges}), only treating Bogoliubov modes now as 
quantum mechanical operators:
\begin{eqnarray}
\hat {\cal Q}^{H/{\cal A}}&=&\int_{I_{B.H}}g^{\mu\nu} \partial_v\delta_{\chi_f}\hat g_{\mu \nu}
\nonumber\\
&=&\int_{I_{B.H}}\partial_v\hat C^{H/{\cal A}}_{zz}D^2_zf\label{quantumcharge}
\end{eqnarray}
As we have discussed, 
classical supertranslations are generating  an infinite family of  black hole metrics. In the same way, the quantum mechanical operator $\hat {\cal Q}^{H/{\cal A}}$
is transforming one particular black hole quantum state into another one:
\begin{equation}
\hat {\cal Q}^{H/{\cal A}}|BH\rangle = |{\widetilde{BH}}\rangle\, .\label{chargeaction}
\end{equation}
Since the black hole state $ |BH\rangle$ is not invariant under the action of  $\hat {\cal Q}^{H/{\cal A}}$, it immediately follows that the supertranslations at the horizon are
spontaneously broken in the black hole vacuum.

\vskip0.2cm

Next we can
expand the Bogoliubov modes in spherical harmonics as 
follows:
\begin{eqnarray}
&~&\delta_{\chi_f}\hat g_{\mu \nu}(v,\theta,\phi)=\sum_{m,l}\delta_{\chi_f}\hat g_{\mu \nu}^{lm}(v,\theta,\phi)\nonumber\\
&=&{1\over \sqrt{\omega_{lm}}}\sum_{m,l}\bigl(\hat b^{\mu\nu}_{ml}Y_{lm}(\theta,\phi)e^{-iv\omega_{ml}}+\hat b^{\dagger,\mu\nu}_{ml}Y_{lm}^*(\theta,\phi)e^{iv\omega_{ml}}\bigr)\, .
\end{eqnarray}
Note that the quantum Bogoliubov modes do depend on $v$ via the non-vanishing frequencies $\omega_{ml}$.
The effective action of the $H/{\cal A}$-modes on the horizon has the following form:
\begin{equation}
S_{eff}=\int dv d\theta d \phi  \bigl(\partial_v\delta_{\chi_f}g_{\mu \nu}(v,\theta,\phi)\bigr)^2
\end{equation}

Then the BMS charge operator $\hat {\cal Q}_{lm}^{H/{\cal A}}$ at the horizon in eq.(\ref{quantumcharge}) that corresponds to  the above quantum Bogoliubov modes takes the following form:
\begin{eqnarray}\label{chargeop}
\hat {\cal Q}_{lm}^{H/{\cal A}}
=-i\sqrt{ \omega_{lm}}\Bigl(\hat b_{ml}e^{-i\omega_{lm}v}-\hat b^{\dagger}_{ml}e^{iv\omega_{ml}}\Bigr)\, .
\end{eqnarray}

 \subsection{Black hole N-portrait and a boundary toy model in one dimension}
 
 In analogy to the N coincident D3-branes of the $AdS_5\times S^5$ geometry, the microscopic picture of 
the Schwarzschild reps. Rindler geometry is proposed to be 
  a bound state (Bose-Einstein condensate) of N graviton particles
 at the quantum critical point.  This model is called the black hole N-portrait \cite{Nportrait,gold}.
  In the black hole quantum N-portrait 
the black hole state $|BH\rangle$ is  given in terms of 
the state of $N$ gravitons at quantum criticality, where $\alpha N = 1$, with $\alpha$ being the gravitational coupling among
the individual graviton modes. 
Discarding the angular $S^2$ part of the Schwarzschild metric or the $R^2$ part of Rindler space,
 we can try to set up a one-dimensional boundary model, which is dual to the two-dimensional Schwarzschild black hole respectively dual to two-dimensional Rindler space.
One  possible toy model for the black hole N-portrait, which features quantum criticality,  is the model of N interacting bosons on a one-dimensional
 ring  \cite{gold}. Further details about this model including its relation to entanglement and quantum information are given in 
 refs.\cite{ring,Dvali:2013vxa,Dvali:2015ywa,Dvali:2015wca,Dvali:2016lnb}.
   This model is often used in condensed matter physics to describe a Bose-Einstein condensate of N 
  bosons.\footnote{This model can be viewed as an alternative to the SYK-model,
 which was argued to be the holographic dual of a two-dimensional AdS black hole.} 
    Note that since the Schwarzschild black hole is  a non-BPS object, we are really dealing with a bound state, created by the  attractive gravitational force among the N graviton
    particles.
   This is in contrast to the BPS D3-branes, which are just coincident, since the gravitational and the p-form forces among the branes
   are opposite and cancel each other.

In this model, 
the quantum  criticality manifests itself in the appearance of classically-gapless Bogoliubov modes that we have identified with the 
classically gapless modes obtained by acting with supertranslations BMS(H/${\cal A}$). 
The energy gap generated by the quantum effects in a given mode of momentum 
 $ {\hbar \over r_S}$  can be estimated as 
\begin{equation}
 \omega_{ml}\sim\Delta E  =  {1\over N} {\hbar \over r_S}\,. 
 \label{boundE}
 \end{equation}

The black hole vacua are then given in terms of coherent states of Bogoliubov modes in the following way:
\begin{equation}
|BH\rangle \equiv |N\rangle=\exp\Bigl( \sum_{lm}\sqrt{n_{lm}}(\hat b_{ml}-\hat b^{\dagger}_{ml})\Bigr)|0\rangle \, ,
\end{equation}
where $n_{lm}$ is the individual occupation numbers for the mode with angular momentum $l,m$.
Then we finally get for the vacuum expectation value of the charge operator (\ref{chargeop})
\begin{eqnarray}
{\cal Q}_{lm}^{H/{\cal A}}=\langle BH|\hat {\cal Q}_{lm}^{H/{\cal A}}|BH\rangle
=\sqrt{\omega_{lm}n_{lm}}\, .
\end{eqnarray}

Therefore,  for a finite energy gap the BMS charges at the horizon are non-zero. 
Using the estimate eq.(\ref{boundE}) we see that the  charges at the horizon scale like
\begin{eqnarray}
{\cal Q}_{lm}\sim{1\over\sqrt{N}}\, .
\end{eqnarray}
Hence for finite $N$, i.e. for finite entropy, the quantum effects provide non-vanishing charges to the black hole. However  in the classical  limit $N\rightarrow\infty$ these charges
vanish, in agreement with the classical  discussion in the previous chapter.

Using eq.(\ref{chargeaction}) we can also the write the vacuum expectation value of the charge operator in the following way:
\begin{eqnarray}
{\cal Q}_{lm}^{H/{\cal A}}=\langle BH|\hat {\cal Q}_{lm}^{H/{\cal A}}|BH\rangle= \langle BH|  {\widetilde{BH}}\rangle    \, .
\end{eqnarray}
It follows that quantum mechanically the overlap between the two states $|BH\rangle$ and $|{\widetilde{BH}}\rangle$ is non-zero, but of oder ${1\over\sqrt{N}}$.
However in the classical limit $N\rightarrow \infty$ the different black hole states are orthogonal to each other.

 \subsection{A quantum resolved bulk metric}

As we have argued,  
 the   classical Goldstone modes are  gapless in the classical limit, i.e.,  they have zero frequencies 
$\omega = 0$. Notice, this  {\it does not} mean that their classical wavelengths $l=\hbar/k\sim r_S$ are infinite
since these modes do not satisfy the dispersion relations of free propagating waves on a flat space-time, but they rather satisfy the dispersion relations in the
Schwarzschild geometry. 
So consider the Schwarzschild  dispersion relation of the  classical Goldstone modes:
\begin{equation}
\hbar\omega=\sqrt{g_{c, vv}k^2}\ , .
\end{equation}
$g_{c, vv}$ is the vv-component of the classical black hole metric
\begin{equation}
g_{c, vv}=(1-\frac{r_S}{r})\,.
\end{equation}
Since $\lim_{r\rightarrow r_S}g_{c, vv}=0$, the above dispersion relation is consistent with $\omega = 0$ and  $1/k\sim r_S$ at the horizon.

Now we go to the quantum case. Here there is a finite energy gap, i.e. the quantum Bogoliubov modes carry an energy of order $1/N$,
whereas the wave length $l$ is still finite.
Now we want to express the relation between $\Delta E$ and $l$ again in terms of a quantum dispersion relation, which now instead of the classical
metric $g_{c, vv}$ contains a quantum contribution to the metric, denoted by $g_{qm, vv}$:
\begin{equation}
\Delta E=\sqrt{g_{qm, vv}k^2}\, .
\end{equation}
Using  $\Delta E = {\hbar \over r_SN}$ and $l=k^{-1}=r_S$, we obtain for the quantum contribution of the metric
\begin{equation}
g_{qm, vv}={\hbar^2\over N^2}\  .
\end{equation}
Combing the classical metric with the quantum contribution, one obtains:
\begin{equation}
g_{vv}=(1-\frac{r_S}{r})+{\hbar^2\over N^2}\  .
\end{equation}
This metric is non-generate.
So we see that quantum corrections of the order ${\hbar^2\over N^2}$ resolve the degeneracy of the classical metric at the horizon, or in other words the horizon gets smeared out and disappears.

\section{Summary}

In this paper we have reviewed and discussed several aspects of supertranslations at the horizon of a Schwarzschild black hole. These horizon supertranslations
form a infinite dimensional, commutative algebra, and they
are accompanied by  Goldstone bosons, which are massless in the classical large N  limit, but acquire masses of order 1/N after including quantum corrections.
Furthermore we have argued that the geometric, semiclassical  large N limit allows for a holographic description, where the holographic boundary of the Schwarzschild geometry
is given by a cone in flat Rindler-space, and where the boundary theory can be described by a large number of interacting bosons. 
It would interesting to see, if there are any relations between the Schwarzschild holography, which is advocated in this paper, and the general space holography, as it is discussed e.g. in   \cite{Bousso:1999cb,Bousso:2002ju}.

Furthermore, it would be also very interesting to more closely relate the considerations and the results of this paper to the scattering amplitudes of two gravitons into N gravitons.
As argued in \cite{Dvali:2014ila} (see also   \cite{Addazi:2016ksu})
the N final state gravitons  in the regime of quantum criticality where $\lambda=\alpha N=1$ can account for the production of a black hole.
As the N gravitons in the final state of the scattering amplitude are not completely soft, but have momenta of the order $\Delta E=1/N$, it is tempting to argue that the N gravitons are closely related
to the quantum Bogoliubov modes considered in this paper; furthermore it seems plausible that  
  the quantum supertranslations
precisely act on these final state gravitons and account in this way for the final black hole entropy. Possible additional massless gravitons, which could be added to the
scattering amplitude, would then correspond to BMS transformations at null infinity, which are however  already eliminated when taking
the quotient of ${\cal A}$-supertranslations.

\section*{Acknowledgements} We like to thank Laura Donney for collaboration on  chapter two of the paper. In also like to thank Gia Dvali for his comments on the manuscript.
Furthermore I like to thank my collaborators Artem Averin, Gia Dvali and Cesar Gomez, with whom a large part of this work was done in collaboration. 
 The work  was supported by  the ERC Advanced Grant 320040 ``Strings and Gravity'', by the DFG interregional research project  
  TRR 33     ``Dark Universe'' and by the DFG Excellence Cluster  ``Origin and Structure of the Universe''.


\newpage



\begin{thebibliography}{99}




\bibitem{Brown:1986nw}
 J.~D.~Brown and M.~Henneaux,
  ``Central Charges in the Canonical Realization of Asymptotic Symmetries: An Example from Three-Dimensional Gravity,''
  Commun.\ Math.\ Phys.\  {\bf 104} (1986) 207.







     \bibitem{bms} H. Bondi, M. G. J. van der Burg, A. W. K. Metzner, ``Gravitational waves in general relativity VII.
Waves from isolated axisymmetric systems", Proc. Roy. Soc. Lond. A 269, 21 (1962);
R. K. Sachs, ``Gravitational waves in general relativity VIII. Waves in asymptotically flat space-time",
Proc. Roy. Soc. Lond. A 270, 103 (1962).


\bibitem{Strominger:2017zoo}
  A.~Strominger,
  ``Lectures on the Infrared Structure of Gravity and Gauge Theory,''
  arXiv:1703.05448 [hep-th].



\bibitem{Strominger:2013jfa}
  A.~Strominger,
  ``On BMS Invariance of Gravitational Scattering,''
  JHEP {\bf 1407} (2014) 152
  [arXiv:1312.2229 [hep-th]].
  
  
  
  T.~He, V.~Lysov, P.~Mitra and A.~Strominger,
  ``BMS supertranslations and Weinberg's soft graviton theorem,''
  JHEP {\bf 1505} (2015) 151
  doi:10.1007/JHEP05(2015)151
  [arXiv:1401.7026 [hep-th]].



 
  F.~Cachazo and A.~Strominger,
  ``Evidence for a New Soft Graviton Theorem,''
  arXiv:1404.4091 [hep-th].
  
  
  D.~Kapec, V.~Lysov, S.~Pasterski and A.~Strominger,
  ``Semiclassical Virasoro symmetry of the quantum gravity $ \mathcal{S}$-matrix,''
  JHEP {\bf 1408} (2014) 058
  doi:10.1007/JHEP08(2014)058
  [arXiv:1406.3312 [hep-th]].
  
  
  A.~Strominger and A.~Zhiboedov,
  ``Gravitational Memory, BMS Supertranslations and Soft Theorems,''
  arXiv:1411.5745 [hep-th].
  






\bibitem{Hawking:2015qqa}
  S.~W.~Hawking,
  ``The Information Paradox for Black Holes,''
  arXiv:1509.01147 [hep-th].
  

  
\bibitem{Hawking:2016msc}
  S.~W.~Hawking, M.~J.~Perry and A.~Strominger,
  ``Soft Hair on Black Holes,''
  arXiv:1601.00921 [hep-th].
  
\bibitem{Hawking:2016sgy}
  S.~W.~Hawking, M.~J.~Perry and A.~Strominger,
  ``Superrotation Charge and Supertranslation Hair on Black Holes,''
  JHEP {\bf 1705} (2017) 161
  doi:10.1007/JHEP05(2017)161
  [arXiv:1611.09175 [hep-th]].
 

\bibitem{Dvali:2015rea}
  G.~Dvali, C.~Gomez and D.~L\"ust,
  ``Classical Limit of Black Hole Quantum N-Portrait and BMS Symmetry,''
  doi:10.1016/j.physletb.2015.11.073
  arXiv:1509.02114 [hep-th].
  
\bibitem{Donnay:2015abr}
  L.~Donnay, G.~Giribet, H.~A.~Gonzalez and M.~Pino,
  ``Super-translations and super-rotations at the horizon,''
  arXiv:1511.08687 [hep-th].



\bibitem{Averin:2016ybl}
  A.~Averin, G.~Dvali, C.~Gomez and D.~L\"ust,
  ``Gravitational Black Hole Hair from Event Horizon Supertranslations,''
  arXiv:1601.03725 [hep-th].
  
  
\bibitem{Averin:2016hhm}
  A.~Averin, G.~Dvali, C.~Gomez and D.~L\"ust,
  ``Goldstone origin of black hole hair from supertranslations and criticality,''
  Mod.\ Phys.\ Lett.\ A {\bf 31} (2016) no.39,  1630045
  doi:10.1142/S0217732316300457
  [arXiv:1606.06260 [hep-th]].


 

	\bibitem{DGGP2} L. Donnay, G. Giribet, H. A. González and M. Pino, ``{Extended Symmetries at the Black Hole Horizon}'',
\href{http://dx.doi.org/10.1007/JHEP09(2016)100}{{\em JHEP} {\bfseries 09} (2016) 100},
\href{https://arxiv.org/abs/1607.05703}{{\ttfamily arXiv:1607.05703 [hep-th]}}.








\bibitem{Banks:2014iha}
  T.~Banks and W.~Fischler,
  ``Holographic Space-time and Newton's Law,''
  arXiv:1310.6052 [hep-th].


  T.~Banks,
  ``Lectures on Holographic Space Time,''
  arXiv:1311.0755 [hep-th].



  T.~Banks,
  ``The Super BMS Algebra, Scattering and Holography,''
  arXiv:1403.3420 [hep-th].
  
  
  
  E.~E.~Flanagan and D.~A.~Nichols,
  ``Conserved charges of the extended Bondi-Metzner-Sachs algebra,''
  arXiv:1510.03386 [hep-th].
  
  
  S.~G.~Avery and B.~U.~W.~Schwab,
  ``BMS, String Theory, and Soft Theorems,''
  arXiv:1506.05789 [hep-th].
  
 
  J.~Ellis, N.~E.~Mavromatos and D.~V.~Nanopoulos,
  ``Information Retention by Stringy Black Holes,''
  arXiv:1511.01825 [hep-th].
  
  
  
  A.~Kehagias and A.~Riotto,
  ``BMS in Cosmology,''
  JCAP {\bf 1605} (2016) no.05,  059
  doi:10.1088/1475-7516/2016/05/059
  [arXiv:1602.02653 [hep-th]].
  
  
  H.~Afshar, S.~Detournay, D.~Grumiller, W.~Merbis, A.~Perez, D.~Tempo and R.~Troncoso,
  ``Soft Heisenberg hair on black holes in three dimensions,''
  Phys.\ Rev.\ D {\bf 93} (2016) no.10,  101503
  doi:10.1103/PhysRevD.93.101503
  [arXiv:1603.04824 [hep-th]].
  
  C.~Eling and Y.~Oz,
  ``On the Membrane Paradigm and Spontaneous Breaking of Horizon BMS Symmetries,''
  arXiv:1605.00183 [hep-th].
  
  
  M.~R.~Setare and H.~Adami,
  ``Near Horizon Symmetries of the Non-Extremal Black Hole Solutions of Generalized Minimal Massive Gravity,''
  arXiv:1606.02273 [hep-th].
  
  
    
  
\bibitem{Blau:2015nee}
  M.~Blau and M.~O'Loughlin,
  ``Horizon Shells and BMS-like Soldering Transformations,''
  arXiv:1512.02858 [hep-th].

 
 
 \bibitem{Afshar:2016uax}
  H.~Afshar, D.~Grumiller and M.~M.~Sheikh-Jabbari,
  Phys.\ Rev.\ D {\bf 96} (2017) no.8,  084032
  doi:10.1103/PhysRevD.96.084032
  [arXiv:1607.00009 [hep-th]].

  M.~M.~Sheikh-Jabbari and H.~Yavartanoo,
  ``Horizon Fluffs: Near Horizon Soft Hairs as Microstates of Generic AdS3 Black Holes,''
  Phys.\ Rev.\ D {\bf 95} (2017) no.4,  044007
  doi:10.1103/PhysRevD.95.044007
  [arXiv:1608.01293 [hep-th]].

  M.~M.~Sheikh-Jabbari and H.~Yavartanoo,
  ``Horizon Fluffs: Near Horizon Soft Hairs as Microstates of Generic AdS3 Black Holes,''
  Phys.\ Rev.\ D {\bf 95} (2017) no.4,  044007
  doi:10.1103/PhysRevD.95.044007
  [arXiv:1608.01293 [hep-th]].
  
  
\bibitem{Setare:2016msj}
  M.~R.~Setare and H.~Adami,
  ``BMS type symmetries at null-infinity and near horizon of non-extremal black holes,''
  Eur.\ Phys.\ J.\ C {\bf 76} (2016) no.12,  687
  doi:10.1140/epjc/s10052-016-4548-0
  [arXiv:1609.05736 [hep-th]].


	\bibitem{Barnich3} G.~Barnich and C.~Troessaert, ``{Aspects of the BMS/CFT correspondence},''
  \href{http://dx.doi.org/10.1007/JHEP05(2010)062}{{\em JHEP} {\bfseries 1005}
  (2010) 062},
\href{http://arxiv.org/abs/1001.1541}{{\ttfamily arXiv:1001.1541 [hep-th]}}.


	\bibitem{Barnich:2001jy} G.~Barnich and F.~Brandt, ``{Covariant theory of asymptotic symmetries,
  conservation laws and central charges},''
  \href{http://dx.doi.org/10.1016/S0550-3213(02)00251-1}{{\em Nucl. Phys.}
  {\bfseries B633} (2002) 3--82},
\href{http://arxiv.org/abs/hep-th/0111246}{{\ttfamily arXiv:hep-th/0111246
  [hep-th]}}.

	\bibitem{Barnich:2007bf} G.~Barnich and G.~Compere, ``Surface charge algebra in gauge theories and thermodynamic integrability,''
  J.\ Math.\ Phys.\  {\bf 49}, 042901 (2008) [arXiv:0708.2378 [gr-qc]].

	\bibitem{Penna:2015gza} R.~F.~Penna, ``BMS invariance and the membrane paradigm,''
   \href{http://dx.doi.org/10.1007/JHEP03(2016)023}{{\em JHEP}{\bf 1603}, 023 (2016)},
 \href{http://arxiv.org/abs/1508.06577}{{\ttfamily arXiv:1508.06577 [hep-th]}}.

\bibitem{Barnich2} G.~Barnich and C.~Troessaert,``Symmetries of asymptotically flat 4 dimensional spacetimes at null infinity revisited,''
 \href{http://dx.doi.org/10.1103/PhysRevLett.105.111103}{{\em Phys.\ Rev.\ Lett.} {\bfseries 105}
  (2010) 111103},
\href{http://arxiv.org/abs/0909.2617}{{\ttfamily arXiv:0909.2617 [gr-qc]}}.

	\bibitem{Barnich4} G.~Barnich and C.~Troessaert,``BMS charge algebra,''
  \href{http://dx.doi.org/10.1007/JHEP12(2011)105}{{\em JHEP} {\bf 1112} (2011) 105} ,
\href{http://arxiv.org/abs/1106.0213}{{\ttfamily arXiv:1106.0213 [hep-th]}}.

	\bibitem{Barnich6} G.~Barnich and C.~Troessaert,
  ``Comments on holographic current algebras and asymptotically flat four dimensional spacetimes at null infinity,''
\href{http://dx.doi.org/ 10.1007/JHEP11(2013)003}{{\em JHEP} {\bf 1311} (2013) 003},
 \href{http://arxiv.org/abs/1309.0794}{{\ttfamily arXiv:1309.0794 [hep-th]}}.
 
 
 
 
 
 	

  
  
	
 
 
\bibitem{Compere:2016gwf}
  G.~Compere,
  ``Bulk supertranslation memories: a concept reshaping the vacua and black holes of general relativity,''
  arXiv:1606.00377 [hep-th].
  
  
\bibitem{Bhattacharjee:2017gkh}
  S.~Bhattacharjee and A.~Bhattacharyya,
  ``Supertranslation and superrotation from soldering transformations,''
  arXiv:1707.01112 [hep-th].


\bibitem{Mirbabayi:2016axw}
  M.~Mirbabayi and M.~Porrati,
  ``Dressed Hard States and Black Hole Soft Hair,''
  Phys.\ Rev.\ Lett.\  {\bf 117} (2016) no.21,  211301
  doi:10.1103/PhysRevLett.117.211301
  [arXiv:1607.03120 [hep-th]].


\bibitem{Bousso:2017dny}
  R.~Bousso and M.~Porrati,
  ``Soft Hair as a Soft Wig,''
  arXiv:1706.00436 [hep-th].
  
  
  
  \bibitem{Bousso:2017rsx}
  R.~Bousso and M.~Porrati,
  ``Observable Supertranslations,''
  arXiv:1706.09280 [hep-th].
 
\bibitem{Bousso:2017xyo}
  R.~Bousso, V.~Chandrasekaran, I.~F.~Halpern and A.~Wall,
  ``Asymptotic Charges Cannot Be Measured in Finite Time,''
  arXiv:1709.08632 [hep-th].


\bibitem{Nportrait}
  G.~Dvali and C.~Gomez,
  ``Black Hole's Quantum N-Portrait,''
  Fortsch.\ Phys.\  {\bf 61} (2013) 742
  [arXiv:1112.3359 [hep-th]];  
  ``Black Hole's 1/N Hair,''
  Phys.\ Lett.\ B {\bf 719} (2013) 419
    [arXiv:1203.6575 [hep-th]].
     ``Quantum Compositeness of Gravity: Black Holes, AdS and Inflation,''
  JCAP {\bf 1401} (2014) 023
  [arXiv:1312.4795 [hep-th]]; 
  ``Black Hole Macro-Quantumness,''
  arXiv:1212.0765 [hep-th]
  
    
  
   
    \bibitem{gold}
  G.~Dvali and C.~Gomez,
  ``Black Holes as Critical Point of Quantum Phase Transition,''
  Eur.\ Phys.\ J.\ C {\bf 74}, 2752 (2014)
    [arXiv:1207.4059 [hep-th]]; 
    



	\bibitem{Moncrief:1983xua} V.~Moncrief and J.~Isenberg, ``Symmetries of cosmological Cauchy horizons,''
\href{http://dx.doi.org/10.1007/BF01214662}{{\em   Commun.\ Math.\ Phys.\ }  {\bf 89}, no. 3, 387 (1983)}.

	\bibitem{Chrusciel} P.~Chru\'{s}ciel, 
  ``The Geometry of Black Holes,'' 
	\href{http://homepage.univie.ac.at/piotr.chrusciel/teaching/Black%20Holes/BlackHolesViennaJanuary2015.pdf}{unpublished notes, available at this URL},
  Erwin Schr\"{o}dinger Institute and Faculty of Physics, 
  University of Vienna (2015).








\bibitem{Maldacena:1997re}
  J.~M.~Maldacena,
  ``The Large N limit of superconformal field theories and supergravity,''
  Int.\ J.\ Theor.\ Phys.\  {\bf 38} (1999) 1113
   [Adv.\ Theor.\ Math.\ Phys.\  {\bf 2} (1998) 231]
  doi:10.1023/A:1026654312961
  [hep-th/9711200].
    
      \bibitem{Bek}
  J.~D.~Bekenstein,
  ``Black holes and entropy,''
  Phys.\ Rev.\ D {\bf 7}, 2333 (1973).



\bibitem{Hawking} 
 S.~W.~Hawking,
  ``Black hole explosions,''
  Nature {\bf 248} (1974) 30.

  
  
  
   \bibitem{ring}
D.~Flassig, A.~Pritzel and N.~Wintergerst,
``Black Holes and Quantumness on Macroscopic Scales,''
  Phys.\ Rev.\ D {\bf 87}, 084007 (2013)
  [arXiv:1212.3344].
  
  
\bibitem{Dvali:2013vxa}
  G.~Dvali, D.~Flassig, C.~Gomez, A.~Pritzel and N.~Wintergerst,
  ``Scrambling in the Black Hole Portrait,''
  Phys.\ Rev.\ D {\bf 88} (2013) no.12,  124041
  doi:10.1103/PhysRevD.88.124041
  [arXiv:1307.3458 [hep-th]].
  
\bibitem{Dvali:2015ywa}
  G.~Dvali, A.~Franca, C.~Gomez and N.~Wintergerst,
  ``Nambu-Goldstone Effective Theory of Information at Quantum Criticality,''
  Phys.\ Rev.\ D {\bf 92} (2015) no.12,  125002
  doi:10.1103/PhysRevD.92.125002
  [arXiv:1507.02948 [hep-th]].
  
\bibitem{Dvali:2015wca}
  G.~Dvali and M.~Panchenko,
  ``Black Hole Type Quantum Computing in Critical Bose-Einstein Systems,''
  arXiv:1507.08952 [hep-th].
  
  
\bibitem{Dvali:2016lnb}
  G.~Dvali, C.~Gomez, D.~L\"ust, Y.~Omar and B.~Richter,
  ``Universality of Black Hole Quantum Computing,''
  Fortsch.\ Phys.\  {\bf 65} (2017) 46
  doi:10.1002/prop.201600111
  [arXiv:1605.01407 [hep-th]].
  

  
  
 
 
 
 
  
 
 

\bibitem{Bousso:1999cb}
  R.~Bousso,
  ``Holography in general space-times,''
  JHEP {\bf 9906} (1999) 028
  doi:10.1088/1126-6708/1999/06/028
  [hep-th/9906022].
  
\bibitem{Bousso:2002ju}
  R.~Bousso,
  ``The Holographic principle,''
  Rev.\ Mod.\ Phys.\  {\bf 74} (2002) 825
  doi:10.1103/RevModPhys.74.825
  [hep-th/0203101].


  
  


\bibitem{Dvali:2014ila} 
  G.~Dvali, C.~Gomez, R.~S.~Isermann, D.~L\"ust and S.~Stieberger,
  ``Black hole formation and classicalization in ultra-Planckian $2\rightarrow N$ scattering,''
  Nucl.\ Phys.\ B {\bf 893}, 187 (2015)
  doi:10.1016/j.nuclphysb.2015.02.004
  [arXiv:1409.7405 [hep-th]].
  
\bibitem{Addazi:2016ksu}
  A.~Addazi, M.~Bianchi and G.~Veneziano,
  ``Glimpses of black hole formation/evaporation in highly inelastic, ultra-planckian string collisions,''
  JHEP {\bf 1702} (2017) 111
  doi:10.1007/JHEP02(2017)111
  [arXiv:1611.03643 [hep-th]].
  



\end{thebibliography}
\end{document}